\documentclass[namedreferences,here]{kluwer}

\newcommand{\ltapprox}{\rlap{\raise.4ex\hbox{$<$}}\lower.5ex\hbox
{{$\sim$}}}
\newcommand{\gtapprox}{\rlap{\raise.4ex\hbox{$>$}}\lower.5ex\hbox
{{$\sim$}}}

\let\oldverbatim\verbatim
\renewcommand{\verbatim}{\expandafter\small\oldverbatim}
\usepackage{epsfig}

\runningtitle{CME WAITING TIMES}

\runningauthor{M.S.\ WHEATLAND}

\begin{document}

\begin{opening}

\title{THE CORONAL MASS EJECTION WAITING-TIME DISTRIBUTION}

\author{M.S. \surname{WHEATLAND}}
\institute{School of Physics, University of Sydney, NSW 2006, Australia
 \protect\\(e-mail: wheat@physics.usyd.edu.au)}


\date{}


\begin{ao}
\end{ao}

\begin{abstract}
The distribution of times $\Delta t$ between coronal mass ejections 
(CMEs) in the Large Angle and Spectrometric Coronagraph (LASCO) CME 
catalog for the years 1996-2001 is examined. The distribution exhibits a 
power-law tail $\propto (\Delta t)^{\gamma}$ with an index 
$\gamma\approx -2.36\pm 0.11$
for large waiting times ($\Delta t>10\,{\rm hours}$). The power-law index
of the waiting-time distribution varies with the solar cycle: for the years 
1996-1998 (a period of low activity), the power-law index is 
$\gamma\approx-1.86\pm 0.14$, 
and for the years 1999-2001 (a period of higher activity), 
the index is $\gamma\approx-2.98\pm 0.20$. 
The observed CME waiting-time distribution, and its variation with the 
cycle, may be understood in terms of CMEs occurring as a time-dependent 
Poisson process.  The CME waiting-time distribution is compared with that 
for greater than C1 class solar flares in the Geostationary Operational 
Environmental Satellite (GOES) catalog for the same years. The flare and 
CME waiting-time distributions exhibit power-law tails with very similar
indices and time variation.
\end{abstract}

\end{opening}

\section{Introduction}

\noindent
Coronal mass ejections (CMEs) are large scale expulsions of plasma and 
magnetic field from the Sun, typically observed in scattered 
white light images of the solar corona and low solar wind. CMEs are 
closely associated with solar flares, explosive events in which magnetic 
energy is released in situ in the solar corona, although the exact 
relationship remains the subject of debate (e.g.\ Kahler, 1992).

The statistics of solar flares have been intensively studied. For example,
hard X-ray studies suggest that the frequency distribution of flare 
energies (the number of flares per unit energy and per unit time) 
follows a power law, $N(E)\propto E^{-1.5}$ \cite{cro&93}. The 
index of the power law does not vary substantially with the solar cycle,
suggesting that it is of fundamental significance, and it was on this basis
that the self-organized critical (SOC) or avalanche model for flares was 
proposed (Lu and Hamilton, 1991; for a review of recent developments of 
this model see Charbonneau et al., 2001).

The statistics of CMEs have also been the subject of scrutiny. Because the 
exact relationship between flares and CMEs remains poorly understood, 
there is fundamental interest in whether flares and CMEs exhibit the same 
statistics. The mass distribution of CMEs appears to follow an exponential
distribution~\cite{jac&how93}, and this may also be the case for the energy
distribution~\cite{jac97}. If this result is confirmed, the difference in 
energy distributions of flares and CMEs may provide a valuable clue to their
respective production mechanisms. Other statistical properties of CMEs that
have been studied include the variation of rate with the solar 
cycle~\cite{web&how94}, the association with different kinds of soft
X-ray flares~\cite{har95}, the possibility of different CME 
classes~\cite{she&99}, the association with sigmoidal structures in active
regions~\cite{can&00}, the association with solar microwave 
bursts~\cite{dou&02}, and the correlation of rate of occurrence with 
properties of originating active regions~\cite{fal&02}.    

Recently another statistic of flares, the distribution of times $\Delta t$
between events (`waiting times'), has attracted attention. 
Boffetta et al.\ (1999)
pointed out that the waiting-time distribution (WTD) for 20 years of 
Geostationary Operational Environmental (GOES) soft X-ray flares exhibits a
power-law tail $\propto (\Delta t)^{\gamma}$ for large waiting times 
(greater than a few hours). They argued that the power law indicates long
correlation times in the flare time series, and attributed particular 
significance to the observed power-law index, $\gamma\approx -2.4$.
They also argued that the appearance of a power law is inconsistent with
the SOC models for flares. 
However, Wheatland and Litvinenko (2002) re-examined the soft X-ray flare
catalog used by Boffetta et al., and found that the power-law 
index of the tail of the WTD varies with the solar cycle, which counts 
against the power law having a fundamental significance. Wheatland and
Litvinenko explained the observed WTD as arising from a Poisson process 
with a time-varying rate. Flares occur essentially independently, i.e.\ 
as a Poisson process in time, but the mean rate of the process varies 
with the solar cycle. Simple theory was presented to account
for the observed form of the WTD in terms of a time-dependent or 
non-stationary Poisson process. In principle this interpretation is 
consistent with SOC models for flares, and Norman et al.\ (2001)  have
demonstrated that an avalanche model with time-dependent driving can
produce a WTD with a power-law tail matching the GOES observations. 
More generally the
appearance of power laws in the waiting-time statistics of SOC models can
be explained in terms of correlated driving~\cite{san&02}. 

In this paper the waiting-time distribution for coronal mass ejections is
examined, based on a catalog of events from the Large Angle and
Spectrometric Coronagraph (LASCO) on the Solar and Heliospheric Observer
(SOHO) spacecraft. To our knowledge this is the first time that waiting
times for CMEs have been examined. One motivation for this study is to 
confirm the results presented in Wheatland and Litvinenko (2002) for soft
X-ray flare waiting times. More generally, as discussed above it is of 
interest to determine whether flares and CMEs share the same statistics,
in this case waiting-time statistics.

The order of presentation of the paper is as follows. In \S\,2.1 the 
catalog on which this study is based is introduced, and the waiting-time
distribution for all events in the catalog is constructed in \S\,2.2. In
\S\,2.3 the time variation of the WTD is examined, by looking at the
distribution for the first and second halves of the catalog. 
In \S\,3 the observed
WTDs are compared with those for soft X-ray flares. In \S\,4 the
observed CME WTDs are explained in terms of simple theory. The theory is
outlined in \S\,4.1, and is then applied to the CME catalog using
a piecewise-constant Bayesian decomposition of the CME time series 
(\S\,4.2). Finally, in \S\,5 the results are discussed and conclusions 
are drawn.

\section{LASCO CME waiting-time distributions}

\noindent

\subsection{LASCO CME catalog}

\noindent
This study is based on the Center for Solar Physics and Space Weather/Naval
Research Laboratory SOHO/LASCO CME catalog (which is available on the
web, at ${\rm http\!:\!\!//cdaw.gsfc.nasa.gov/CME\_list/}$). The catalog
lists all CMEs identified by LASCO operators from the start of SOHO 
observations in January 1996 to the end of 2001, a total of 4645 events.
For each event the catalog lists the date and time of the first 
appearance of the CME in the LASCO C2 coronagraph field of view, the 
central position angle of the CME, estimates of the CME speed, and a number
of other details. In this study only the date and time of first appearance 
in the C2 coronagraph is used.

Figure~1 shows monthly numbers of CMEs in the catalog (solid histogram),
together with the monthly sunspot numbers (dotted histogram).\footnote{The
sunspot numbers are available from the National Geophysical Data Center at
${\rm ftp\!\!:\!\!//ftp.ngdc.noaa.gov/STP/SOLAR\_DATA}$.} 
The CME numbers closely track the sunspot number, as 
pointed out by~\inlinecite{web&how94}. 
Figure~1 also indicates by vertical dashed 
lines the start and end of two major data gaps, that were due to the 
temporary loss of the spacecraft in the second half of 1998, and to a 
second interruption to the mission late in the year. In this study these 
gaps are taken into account.  

\begin{figure}[h]
\vspace{0.5cm}
\centerline{\epsfig{file=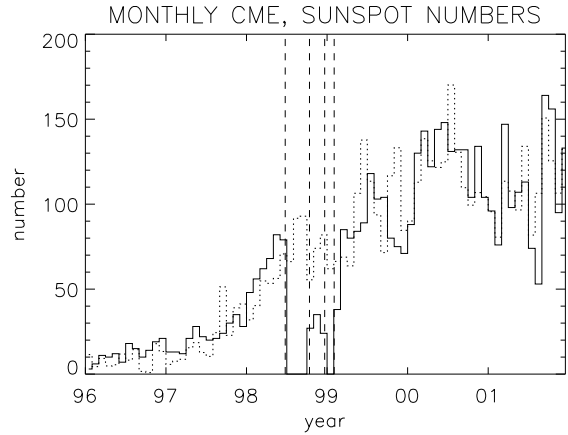,width=20pc}}
\caption{ \label{fig:f1}Monthly numbers of CMEs in the LASCO catalog for
1996-2001 (solid histogram) and monthly sunspot numbers for the same period
(dotted histogram). The dashed vertical lines indicate the beginning and 
end of two major data gaps.}
\end{figure}   

\subsection{WTD for 1996-2001}

\noindent
Figure~2 shows the waiting-time distribution constructed for all events in
the SOHO/LASCO CME catalog, presented as a log-log plot. The distribution
exhibits a clear power-law tail $\propto (\Delta t)^{\gamma}$ for waiting 
times $\Delta t$ greater than about 10 hours, and a fit to the distribution 
(for $\Delta t>10\,{\rm hours}$) gives $\gamma\approx -2.36\pm 0.11$. The
power-law fit is shown by the solid line in the figure. 

Figure~2 confirms that the WTD exhibits similar power-law behavior to that 
observed for the GOES soft X-ray flares~\cite{bof&99,whe&lit02}. A detailed
comparison with the flare data is made in \S\,3.  

For short waiting times (less than a few hours) the WTD in Figure~2 
exhibits widely
varying values in adjacent bins. This effect is most likely due to the
discrete times between C2 observations. Typical observing sequences for the
LASCO instrument lead to certain intervals between C2 observations being
more common, which influences whether CMEs that are closely spaced in time 
are observed. 

\begin{figure}[h]
\vspace{0.5cm}
\centerline{\epsfig{file=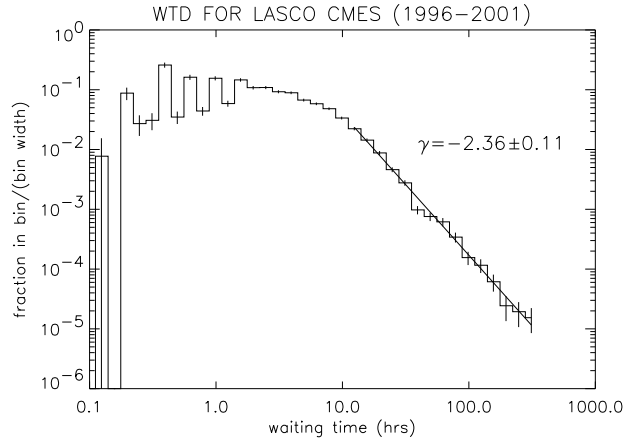,width=20pc}}
\caption{ \label{fig:f2}Waiting-time distribution for LASCO CMEs for the 
years 1996-2001.}
\end{figure}   

\subsection{Time variation of WTD}

\noindent
Figure~1 shows that the observing period (1996-2001) encompasses an initial
period of solar minimum, followed by the rise and part of the maximum of 
a solar cycle (cycle 23), as defined by sunspot number. To examine the secular 
variation of the WTD, we have divided the dataset into two halves: the 
period 1996-1998, and the period 1999-2001. The first half includes solar 
minimum and the rise to maximum, and the second half is essentially a 
period of maximum activity.

Figure~3 shows the WTDs for these two periods. The power-law tail of the
WTD clearly varies with time: for 1996-1998 the index of the power law is
$\gamma\approx -1.86\pm 0.14$, and for 1999-2001 the index is
$\gamma\approx -2.98\pm 0.20$. The indices were again determined by fits to
the observed distributions for $\Delta t>10\,{\rm hours}$, and the fitting
power laws are shown in the figure by the solid lines.

\begin{figure}[h]
\vspace{0.5cm}
\centerline{\epsfig{file=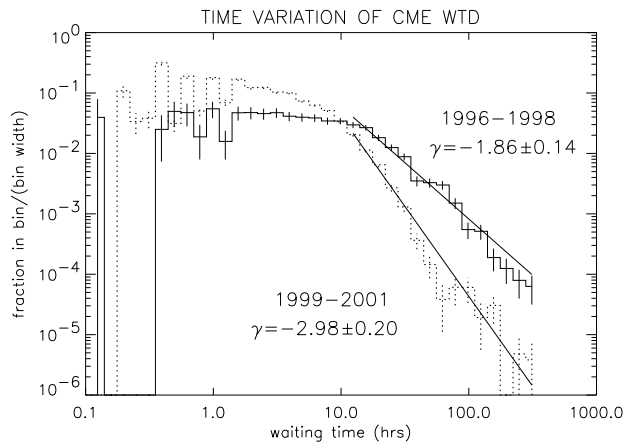,width=20pc}}
\caption{ \label{fig:f3}Waiting-time distribution for LASCO CMEs for the
years 1996-1998 (solid), and for the years 1999-2001 (dotted).}
\end{figure}   
\noindent

It is easy to understand, at a qualitative level, the origin of the 
variation shown in Figure~3. 
During the period of higher activity (1999-2001) the mean rate at
which CMEs occur is higher, as shown by the monthly numbers in Figure~1.
The average waiting time is the reciprocal of the mean rate, so the
average waiting time is shorter during 1999-2001, i.e.\ there are
fewer large waiting times. However, this explanation does
not account for the observed functional forms of the WTDs. A detailed 
quantitative explanation for the observed behavior is outlined in \S\,4.

\section{Comparison with flare WTDs} 

\noindent
It is interesting to compare the CME WTDs with those for soft X-ray flares 
for the same periods of time. For the flare data we use the GOES soft X-ray
flare catalog, which spans the years 1975-2002.\footnote{The GOES catalog 
is available online at 
${\rm ftp\!\!:\!\!//ftp.ngdc.noaa.gov/STP/SOLAR\_DATA}$} 
The events in the catalog were selected from soft X-ray time histories 
measured by the GOES satellites. The soft X-ray background varies by more 
than an order of magnitude over the solar cycle. Consequently small flares 
are not observed at solar maximum owing to the increased background, an 
effect that biases WTDs constructed from all events in the catalog (for 
a more extensive discussion of this point, see Wheatland and Litvinenko, 
2002). 
To correct for this effect we have restricted the GOES events to those 
exceeding C1 class (a peak flux greater than 
$10^{-6}\,{\rm W}\,{\rm m}^{-2}$ at the spacecraft). This flux is an
appropriate threshold because it is typical of background values at solar 
maximum. 
Note that for the CMEs, no similar `thresholding' is applied,
because the detection of events is essentially limited by the sensitivity
of the instrument, and not by selection against a time-varying background. 
An exception to this is that at times of higher activity multiple CMEs 
occurring nearly simultaneously are more common, and in this situation
some events may be missed due to a `background' of other events. 
However, this effect only influences the number of short waiting times,
and not the tail of the WTD.

Figure~4 shows the WTDs for the CME catalog (solid histogram), and for the
GOES soft X-ray flares of greater than C1 class (dotted histogram) for the
duration of the CME catalog, i.e.\ for the 
years 1996-2001. Both distributions exhibit similar power-law tails. Fits
to the behavior of each distribution for $\Delta t>10\,{\rm hours}$ give 
power-law indices $\gamma\approx -2.36\pm 0.11$ for the CME
distribution, and $\gamma\approx -2.26\pm 0.11$ for the flare distribution.
Hence the two indices agree with one another within the stated
uncertainties. 
 
\begin{figure}[h]
\vspace{0.5cm}
\centerline{\epsfig{file=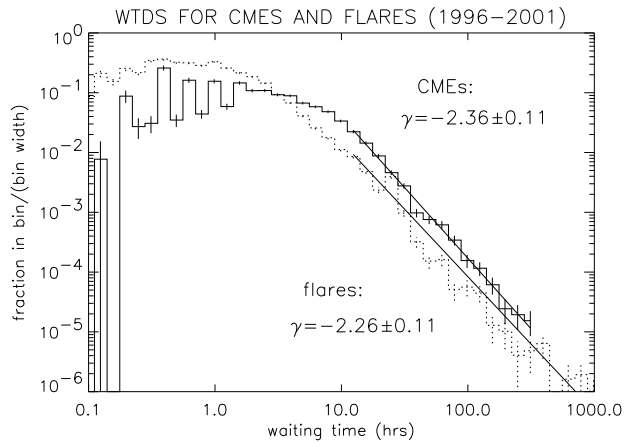,width=20pc}}
\caption{ \label{fig:f4}Waiting-time distribution for LASCO CMEs for the
years 1996-2001 (solid histogram), and for GOES soft X-ray flares of
greater than C1 class for the same years (dotted histogram).}
\end{figure}   

As shown in Figure~3, the power-law tail in the CME WTD is different for the
periods 1996-1998 and 1999-2001. It was shown in Wheatland and Litvinenko
(2002) that the WTD for the GOES events varies with the solar cycle, but
how does the time variation of the flare distribution compare with that
for the CME WTD? Figure~5 shows the CME and flare WTDs for the years
1996-1998 (top panel) and for the years 1999-2001 (bottom panel). In each
panel the CME distribution is shown as a solid histogram and the flare 
distribution is shown as a dotted histogram. Power-law fits to the tail of
each distribution ($\Delta t>10\,{\rm hours}$) are indicated by solid lines,
and by the numerical values of indices shown in the figure. For the years
1996-1998 the CME index is $\gamma\approx -1.86\pm 0.14$, and the flare
index is $\gamma\approx -1.75\pm 0.08$. For the years 1999-2001 the CME
index is $\gamma\approx -2.98\pm 0.2$, and the flare index is
$\gamma\approx -3.04\pm 0.19$. Hence we find that the two phenomena
exhibit power-law indices that vary in essentially the same way with time.
The observed power-law indices are the same within uncertainties.

\begin{figure}[h]
\vspace{0.5cm}
\centerline{\epsfig{file=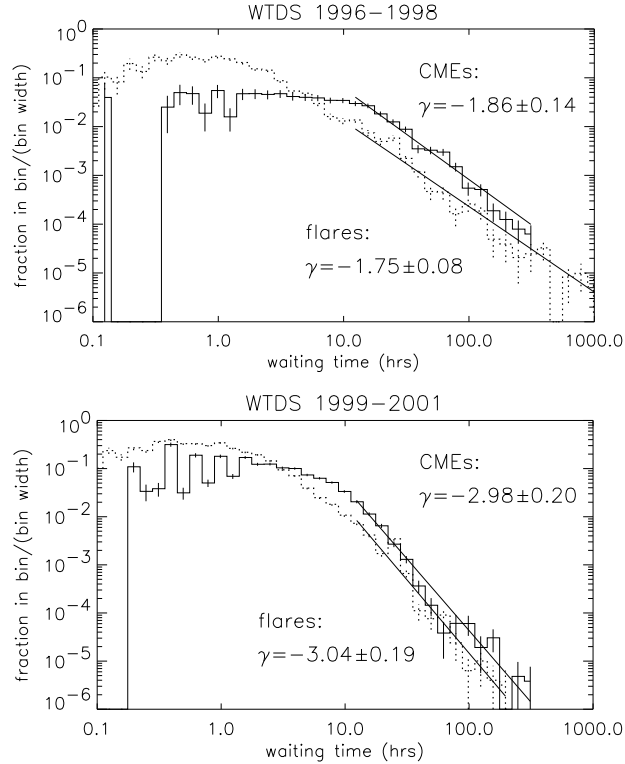,width=20pc}}
\caption{ \label{fig:f5}Top panel: WTDs for LASCO CMEs (solid histogram) 
and for GOES soft X-ray flares (dotted histogram) for the years 1996-1998.
Bottom panel: The same, for the years 1999-2001.}
\end{figure}   

\section{Explaining the observed WTDs} 

\subsection{Theory} 
\noindent
The observed waiting-time distributions for LASCO CMEs may be understood
in terms of CMEs occurring as a time-dependent Poisson process, i.e.\ as a
sequence of independent events, with a mean rate that varies with time. 
The necessary theory was presented in Wheatland and Litvinenko (2002), 
and is summarized here.

A Poisson process is a sequence of events in time such that there is a
constant probability per unit time $\lambda$ of an event occurring. The
quantity $\lambda$ is referred to as the rate of the process. The
waiting-time distribution for a Poisson process is a simple exponential,
\begin{equation}\label{eq:pois1}
P(\Delta t)=\lambda e^{-\lambda \Delta t}.
\end{equation}

A sequence of independent events with a mean rate $\lambda=\lambda(t)$ that
varies with time $t$ defines a time-dependent Poisson process. Provided the
rate varies slowly (with respect to the average waiting time), the WTD for
a time-dependent Poisson process may be written
\begin{equation}\label{eq:app_avnonH}
P(\Delta t)=\frac{1}{N}\int_0^T\lambda (t)^2e^{-\lambda (t) \Delta t}dt,
\end{equation}
where $N$ is the total number of observed events and $T$ is the total
observing time. Equation~(\ref{eq:app_avnonH}) may be 
recast in the form
\begin{equation}\label{eq:lap}
P(\Delta t)=\frac{1}{\overline \lambda}\int_{0}^{\infty}\lambda^2f(\lambda)
 e^{-\lambda\Delta t}\,d\lambda,
\end{equation}
where $f(\lambda)$ is the time distribution of the rate, i.e.\
$f(\lambda)d\lambda$ is the fraction of time that the rate is in the 
range $(\lambda,\lambda+d\lambda)$, and where
\begin{equation}
\overline{\lambda}=\frac{N}{T}=\int_{0}^{\infty}\lambda
f(\lambda)\,d\lambda
\end{equation}
is the average rate.

For the special case of a piecewise-constant Poisson process, i.e.\ a
Poisson process consisting of a series of constant rates $\lambda_i$ for
intervals $t_i$, Equation~(\ref{eq:app_avnonH}) becomes
\begin{equation}\label{eq:app_discr}
P(\Delta t)= \frac{1}{\overline \lambda}\sum_i\frac{t_i}{T}\lambda_i^2
  e^{-\lambda_i\Delta t}.
\end{equation}

To understand the behavior of the tail of the WTD, note
that Equation~(\ref{eq:lap}) implies that the asymptotic ($\Delta
t\rightarrow\infty$) behavior of $P(\Delta t)$ depends on the behavior of
$\lambda^2f(\lambda)$ for small values of $\lambda$. Specifically if
$f(\lambda)$ has a power-law form $f(\lambda)\propto \lambda^{\alpha}$ for
$\lambda\rightarrow 0$, then we have 
$P(\Delta t)\propto (\Delta t)^{-(3+\alpha)}$ for $\Delta t\rightarrow\infty$.
This result holds for $\alpha>-3$, and in particular includes the case 
$\alpha=0$, when $f(\lambda)$ is a constant for small $\lambda$. Hence
a power-law tail in the WTD arises from power-law behavior in the time
distribution of the rate, for low rates (including the case when the 
distribution is flat for low rates).

\subsection{Comparison of observations with the piecewise-constant model} 

\noindent
To apply the theory outlined in \S\,4.1, it is necessary to estimate the
time variation of the rate of CME occurrence, $\lambda=\lambda (t)$, for
events in the LASCO catalog. Note that there is no doubt that the CME 
rate is time varying, as evidenced by the secular change in the monthly 
values in Figure~1.

To approximate the rate we apply the Bayesian Blocks procedure~\cite{sca98},
which takes the time history of CME events and decomposes it into a sequence
of piecewise-constant rates $\lambda_i$ and intervals $t_i$. This procedure
was previously applied to the GOES flares~\cite{whe00,whe&lit02}. The
result of applying the Bayesian Blocks procedure to the LASCO catalog is
shown in Figure~6. The top panel shows the cumulative number of CMEs in
the catalog as a function of time --- the gradient of this
graph indicates the mean rate of occurrence of CMEs. The two major data 
gaps (discussed in \S\,2.1) are shown by the vertical dashed lines. 
The bottom panel of Figure~6 shows the Bayesian rates and intervals. The
Bayesian procedure has decomposed the six years of observations into 33
intervals during which the rate was approximately constant. 

\begin{figure}[h]
\vspace{0.5cm}
\centerline{\epsfig{file=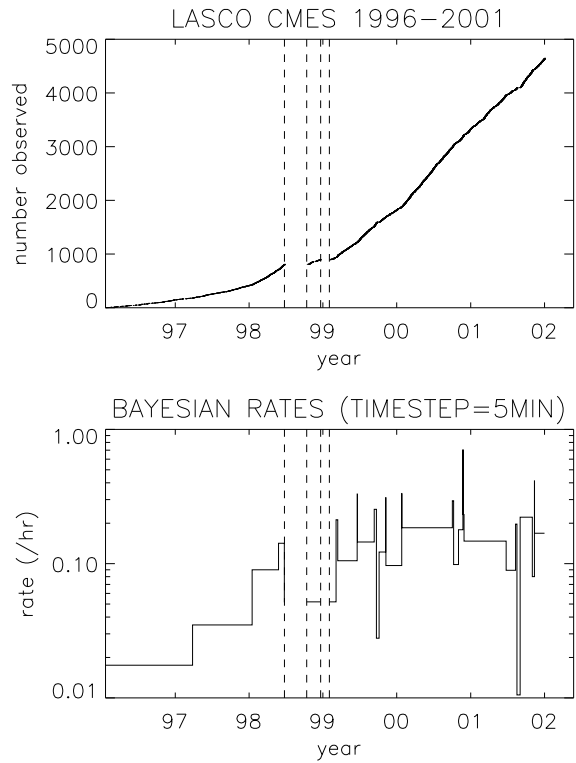,width=20pc}}
\caption{ \label{fig:f6}Top panel: Cumulative number of CMEs in the LASCO 
catalog versus time. Bottom panel: Bayesian estimate of piecewise-constant
rates and intervals for the LASCO CMEs. In both panels the dashed vertical
lines indicate the major data gaps.}
\end{figure}   

The Bayesian procedure is essentially autonomous, i.e.\ makes its own 
decisions during the rate decomposition shown in the bottom panel of
Figure~6. However, a number of free parameters need to be
specified. First, the method requires observations in discrete timesteps.
For simplicity we have presented the data in timesteps of five minutes.
Second, the method involves a prior-odds ratio, which has been set to two.
This means that the data for a given interval must be considered to be 
twice as likely to be produced by a two-rate process by comparison with a 
single-rate process for the two-rate process to be adopted as the model. 
Third, the minimum number of events in an interval is chosen to be three.
In practise we find that the rate decomposition is relatively insensitive
to the choices of these parameters.

Based on the rates and intervals determined by the Bayesian procedure we 
can construct a model piecewise-constant Poisson WTD, using 
Equation~(\ref{eq:app_discr}). The result is shown in Figure~7 by the solid
curve. The histogram is the observed WTD for all of the LASCO events, as
also shown in Figures~2 and~4.
Figure~7 demonstrates that the piecewise-constant model provides a 
reasonable representation of the observed WTD, including reproducing the 
extended tail of the distribution. It should be noted that the 
piecewise-constant model involves a fairly crude approximation --- the
true rate of occurrence of CMEs is expected to be continuously varying. 
Despite this limitation, and the approximate method of rate determination, 
the piecewise-constant Poisson model is seen to reproduce the observed 
form of the WTD. 

\begin{figure}[h]
\vspace{0.5cm}
\centerline{\epsfig{file=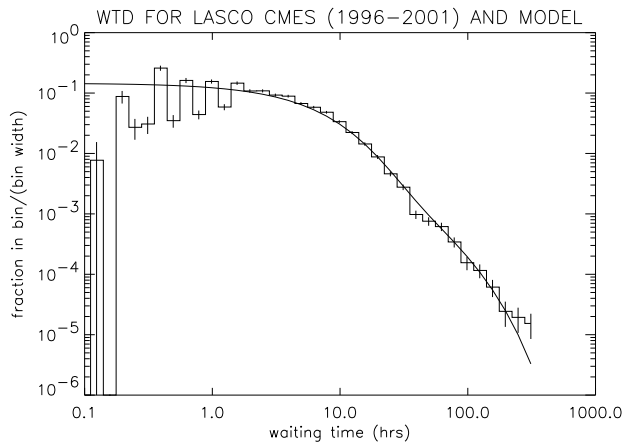,width=20pc}}
\caption{ \label{fig:f7} WTD for the LASCO CMEs for 1996-2001 (histogram),
together with the WTD for the piecewise-constant Poisson model, with rates
and intervals taken from the Bayesian estimates in the bottom panel of
Figure~6.}
\end{figure}   

It is straightforward to construct the piecewise-constant 
model~(\ref{eq:app_discr}) for the periods 1996-1998 and 1999-2001 
separately, based 
on the Bayesian estimates in the lower panel of Figure~6. The result is
shown in Figure~8. The histograms show the WTDs for the two periods, 
as also shown in Figure~3. The curves show the piecewise-constant 
models for each period. The piecewise-constant models are seen to
qualitatively reproduce the observed WTDs. Once again, it 
should be noted that although the models are crude, the observations are
reproduced reasonably well.

\begin{figure}[h]
\vspace{0.5cm}
\centerline{\epsfig{file=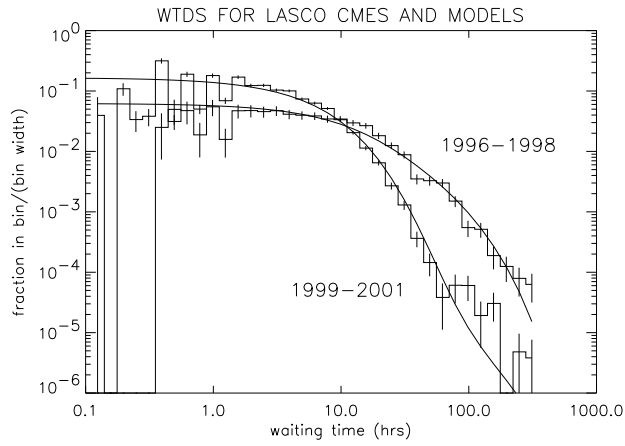,width=20pc}}
\caption{ \label{fig:f8} Observed LASCO CME WTDs for 1996-1998 and for
1999-2001 (solid histograms), together with piecewise-constant Poisson 
models (solid curves). }
\end{figure}

\section{Conclusions}

\noindent
In this paper the waiting-time distribution for coronal mass ejections 
is examined, based on the LASCO CME catalog, which spans the years
1996-2001. The main results are first that the WTD for the entire catalog 
exhibits a power-law
tail (for waiting times $\Delta t$ larger than about 10 hours), i.e.\ the
distribution has the form $\propto (\Delta t)^{\gamma}$, with a power-law
index 
$\gamma\approx -2.36\pm 0.11$. Second, the power-law index of the tail
varies with time, so that the power law is significantly steeper at times 
of higher activity, i.e.\ when CMEs are more frequent. 
Third, the power laws in CME WTDs are 
the same as those found in the WTDs of GOES soft X-ray flares, for the same
years (i.e.\ flares and CMEs exhibit the same power-law tails in their 
WTDs, and the same time variation of those tails).
Finally, the observed WTDs can be explained in terms of CMEs occurring as 
a Poisson process in time, with a time-varying rate. 

The theory presented in \S\,4.1 describes the observed WTDs as arising from
a Poisson process with a time-dependent rate $\lambda=\lambda (t)$. This
means that the theory assumes that CMEs occur independently of one another,
and with a probability per unit time that varies with time. It is necessary 
to explain both of these aspects of the model. Concerning the independence
(randomness) of CME occurrence, the first point to note is that consecutive
CMEs in the catalog may often originate from different regions on the Sun, 
which are expected to produce CMEs independently. However, at times of
solar minimum the majority of CMEs may originate from one or a few active
regions at any given time. Although the detailed theory of the CME 
mechanism remains to be 
worked out, numerical models often involve complex magnetic configurations 
that are presumably reached via complicated processes of coronal magnetic 
field evolution (e.g.\ Amari et al., 2000). Such processes may explain the 
origin of the apparent independence of successive CMEs from a common 
region on the Sun. 
Regarding the time-varying rate of CME occurrence, on short timescales the
CME rate is expected to vary due to the appearance and disappearance of 
active regions with suitably complex magnetic configurations. On longer 
timescales the rate is expected to vary with the solar cycle due to the 
cyclic variation in the number of active regions (see Figure~1).

As noted in the introduction, the poor understanding of the relationship 
between flares and CMEs means that there is interest in whether 
flares and CMEs exhibit the same statistics. In \S\,3 of this paper it is
shown that flare and CME waiting-time statistics are very similar, and in 
particular the two phenomena exhibit the same power-law tails in their
WTDs. The theory in \S\,4.1 of this paper shows that, assuming the 
occurrence of flares and CMEs can be understood as a time-dependent Poisson
process, the observed WTD depends only on the time distribution of the
rate of events. In this case the appearance of similar WTDs implies that
both flares and CMEs occur with a rate that varies with time in a similar 
way. At face value this result suggests a common origin for flares and 
CMEs. However, it should be noted that there are pieces of observational 
evidence suggesting distinct origins for flares and CMEs. For example,
at solar maximum a significant number of CMEs originate from very high
solar latitudes, well away from active regions~\cite{hun99}.
 
The appearance of different power-law tails in the CME WTDs was explained
in \S\,4.1 in terms of power-law like behavior in the time distribution 
of occurrence rate at low rates. At times of high activity the observed
WTD has an index close to $\gamma=-3$, which implies that the 
at these times the time-distribution of rates is approximately flat at 
low rates. For times of lower activity the WTD has a shallower power law, 
implying that at these times there is a decreasing power law in the time 
distribution of rates at low rates. This behavior is consistent with that
previously found for flares~\cite{whe&lit02}. The origin of this 
interesting difference in CME (flare) behavior between different phases of
the cycle remains an open question. 

The absence of a fixed power law in the tail of the CME and flare WTDs 
suggests that the power law does not have fundamental significance, 
such as that attributed to the power law in the flare energy distribution. 
The theory outlined in \S\,4.1 explains the observed power laws, and 
suggests that they will appear generically for phenomena involving 
independent events with a time-varying rate. 
However, the power laws do have some significance, in that they contain 
important information about the rates at which CMEs and flares occur. 
In particular, the implication of power laws in the underlying rate 
distributions may be a valuable clue for understanding the mechanisms of 
energy storage and release in flares.

\section*{Acknowledgements}

\noindent 
M.S.W. acknowledges the support of an Australian Research Council QEII 
Fellowship, and thanks an anonymous referee whose comments helped to 
improve this paper. The CME catalog used in this work was generated by the Center 
for Solar Physics and Space Weather, The Catholic University of America 
in cooperation with the Naval Research Laboratory and NASA. 
SOHO is a project of international cooperation between ESA and NASA.

\end{document}